\documentclass[3p]{elsarticle}

\usepackage{setspace}
\usepackage[hidelinks]{hyperref}
\usepackage{amsmath, amsthm, amssymb}
\usepackage{mathrsfs}
\usepackage{xcolor}

\def\sym#1{\ifmmode^{#1}\else\(^{#1}\)\fi}

\biboptions{authoryear, sort}

\begin{document}

\begin{frontmatter}

\title{The citation disadvantage of clinical research}

\author{Qing Ke\corref{corrauthor}}
\address{Center for Complex Network Research, Northeastern University, Boston, MA 02115, USA}
\cortext[corrauthor]{Corresponding author}
\ead{q.ke@northeastern.edu}

\begin{abstract}
Biomedical research encompasses diverse types of activities, from basic science (``bench'') to clinical medicine (``bedside'') to bench-to-bedside translational research. It, however, remains unclear whether different types of research receive citations at varying rates. Here we aim to answer this question by using a newly proposed paper-level indicator that quantifies the extent to which a paper is basic science or clinical medicine. Applying this measure to 5 million biomedical papers, we find a systematic citation disadvantage of clinical oriented papers; they tend to garner far fewer citations and are less likely to be hit works than papers oriented towards basic science. At the same time, clinical research has a higher variance in its citation. We also find that the citation difference between basic and clinical research decreases, yet still persists, if longer citation-window is used. Given the increasing adoption of short-term, citation-based bibliometric indicators in funding decisions, the under-cited effect of clinical research may provide disincentives for bio-researchers to venture into the translation of basic scientific discoveries into clinical applications, thus providing explanations of reasons behind the existence of the gap between basic and clinical research that is commented as ``valley of death'' and the commentary of ``extinction'' risk of translational researchers. Our work may provide insights to policy-makers on how to evaluate different types of biomedical research.
\end{abstract}

\begin{keyword}
biomedicine \sep basic research \sep clinical medicine \sep citation analysis \sep bias
\end{keyword}

\end{frontmatter}

\section{Introduction}

Citation-based indicators, such as h-index \citep{Hirsch-index-2005} and Journal Impact Factor \citep{Garfield-jif-2006}, have been increasingly employed in the evaluation of research performance, playing a significant role in various decision-making processes, like funding, hiring, and promotion, and in the ranking of diverse actors in science. Such practice, however, has received criticisms \citep{MacRoberts-problem-1989, Seglen-why-1997, Nature-impact-2005}, and a major argument behind those criticisms is that citation data suffers from the presence of several biases. For example, it has been shown that citations have been inflating over time \citep{Persson-inflate-2004}, so that later papers on average are cited more than earlier ones. Another bias is that there is a significant variation of citations for papers in different research fields \citep{Bornmann-what-2008}; e.g., biology papers on average receive a larger number of citations than mathematics papers. Recently, \citet{Wang-bias-2017} found that there is also bias against novel researches---papers that make first-ever combinations of existing knowledge components; they are more likely to be hit works, but display delayed recognition and are published in journals with low Impact Factor (IF). Given the widespread usage of citation-based indicators, it is important to understand what other potential biases are and carry out citation analysis in a less biased way.

In this article, we focus on biomedical research and investigate whether and to what extent another bias, which we call research level bias, is present in citation data. Biomedicine is a broad area of science that studies biological process of life and the cause and treatment of disease. It is characterized by diverse types of research activities, ranging from basic science that advances the understanding of living organisms by announcing scientific discoveries, to clinical research that studies the treatment of disease, to translational research that aims for translating basic scientific discoveries into the development of useful clinical applications that benefit patients. The questions we seek to answer here are: Is there systematic difference in citations for papers belonging to different types of researches? Does certain type of biomedical research exhibit particular disadvantages in receiving citations? These questions are important because the existence of the research level bias in citations is of particular concerns for funding agencies and institutions, given that they have increasingly relied on citation-based bibliometric indicators for research evaluation. To the best of our knowledge, our work is the first to answer these questions systematically. We also remark that this potential research level bias goes beyond the well-known and widely discussed field bias, because one field may encompass more than one type of researches. For instance, within oncology, a well-operationalized field\footnote{In the Web of Science database, there is a Subject Category label called ``Oncology.''}, papers can be basic science about the identification of tumor suppressing genes, or clinical research about the development of cancer drugs.

Here, drawing on more than 5 million biomedical research papers indexed in the MEDLINE database and published in a 23-year period (1980--2002), we find that there is disadvantage in getting citations for clinical research papers. Our analysis relies on a newly minted paper-level indicator that quantifies the extent to which a biomedical paper is oriented towards basic science or clinical research. Our results suggest that papers that are oriented towards clinical research tend to garner far fewer citations than papers oriented towards basic science. Clinical papers also have a much lower chance for being ranked in the top 1\% most cited papers among those in the same field and the same year. Our further analysis using short-, medium-, and long-term citation windows reveals that this citation disadvantage weakens, but still persists, as we increase window length.

To the extent of the reported citation disadvantage of clinical research is a long-standing concern about the existence of the gap between basic biomedical research and clinical research. It has been argued that tremendous progresses made in basic science do not give rise to commensurate, useful clinical applications that benefit patients \citep{Butler-valley-2008}. In the attempt to bridge this gap, funding bodies have been increasingly emphasized the demonstration of societal impact (e.g., impact on patients) of research and made ongoing efforts to initialize large translational research programs like Clinical and Translational Science Awards \citep{Zerhouni-roadmap-2003}. Translation, however, is a decade-long process with frequent failures \citep{Contopoulos-life-2008, Morris-answer-2011}, which makes the academic performance evaluation a rather challenging task. Convenient, citation-based metrics, such as the number of publications in high IF journals and h-index, are instead adopted for the task. This, however, may provide disincentives for basic researchers to venture into the translation of basic science results into clinical products, as clinical papers receive fewer citations, which in turn are emphasized in performance evaluation. The citation disadvantage of clinical research, therefore, may help explain and further reinforce the current situation in the biomedical research enterprise, where the gap has been described as the ``valley of death'' \citep{Butler-valley-2008, Pasterkamp-lost-2016} and ``translational researchers are at risk of becoming `extinct'" \citep{Pasterkamp-lost-2016}. From a policy perspective, our findings may provide insights to policy-makers on how to evaluate different types of biomedical research.

\section{Literature review} \label{sec:lit}

\subsection{Basic and clinical research}

The notions of basic and clinical research have long been adopted by diverse fields to characterize various aspects of the biomedical research enterprise. \citet{Moses-financial-2005}, for example, examined U.S. funding trends of basic and clinical research received from government, foundations, industry and other sources. \citet{Zinner-life-2009} conducted a mailed survey of faculty in U.S. academic medical centers, asking about type of research (basic, clinical trials, epidemiology etc.) as well as funding, publishing, and patenting behavior. They found that 33.6\% of faculty exclusively performed basic research, in contrast to 9.1\% translational researchers. \citet{Levitt-future-2017} looked at the age structure of basic science faculty who are eligible for grant and found a steady decrease of the number of such faculty. Their analysis further indicates that the fraction of faculty who received R01 grant from the U.S. National Institutes of Health (NIH) decreased for younger faculty and increased for older ones. Many studies set out to understand the life-cycle of translational science and showed that this process takes many years and fails frequently \citep{ Contopoulos-life-2008, Khoury-emerge-2010, Morris-answer-2011, Collins-reengineer-2011, Williams-from-2015}. Finally, many scholars have argued for the public support of basic science \citep{Lane-measuring-2011, Moses-biomed-2011, Press-special-2013}, and several prior works in the innovation literature studied the returns of investments in science and the interactions between basic scientific advances and practical technological progress \citep{Toole-impact-2012, Li-applied-2017}.

\subsection{Indicators for research level}

Given the widespread use of the two terms, how can one operationalize them at the individual paper level? How to infer whether a given paper is basic science or clinical research? Such an operationalization is especially important for science and technology studies, because the availability of an indicator for research level opens venues for systematic investigations of the aforementioned characterizations of biomedicine. In the present work, we study how the research level of a paper is associated with the number of citations it receives.

The earliest study that proposed an indicator for research level may be \citet{Narin-structure-1976}, where they subjectively classified journals into four levels of biomedical research. They are, in the order from most basic to most clinical, basic research (L4), clinical investigation (L3), clinical mix (L2), and clinical observation (L1). For each level, they further selected one prototypical journal: \emph{Journal of Biological Chemistry} for L4, \emph{Journal of Clinical Investigation} for L3, \emph{New England Journal of Medicine} for L2, and \emph{Journal of the American Medical Association} for L1. All papers published in a journal then carry the research level of the journal. This four-level categorization has later been used by Narin and his colleagues in studies about patent-to-paper citations, showing that papers that are cited by patents are mostly from L4 and therefore arguing the advantage of basic research in generating technological innovations \citep{Narin-linkage-1997, McMillan-biotech-2000}. The classification schema has also been employed in research evaluation \citep{Lewison-effect-1998, Lewison-biblio-1999}.

Recent proposals of research level indicators focused on papers rather than journals and leveraged Medical Subject Heading (MeSH) to quantify research level. MeSH terms are hierarchically-organized controlled medical vocabularies. They are maintained and used by the U.S. National Library of Medicine (NLM) to index and retrieve MEDLINE papers. The MeSH terms of a paper describe the content of the research, and the assignment of terms to papers are performed not by authors but rather by trained staffs at NLM, to guarantee the consistence of the assignment.

The main idea behind using MeSH terms to infer research level is that certain organism-related terms are for labeling the type of  subjects studied, which corresponds to the key element in the commonly-adopted operational definitions of basic and clinical research. Basic science, for example, has been operationalized as researches with experiments performed at the level of cell and non-human animals. Those researches are conducted without a specific application objective (e.g., developing a new drug), but rather to make scientific discoveries\footnote{Curiosity Creates Cures: The Value and Impact of Basic Research; \url{https://www.nigms.nih.gov/Education/Pages/factsheet_CuriosityCreatesCures.aspx}}. Basic research often uses model organism like \emph{Caenorhabditis elegans} as experiment subject. The appearance of cell and animal MeSH terms in a paper therefore is indicative that it is basic science. Clinical research, on the other hand, has been defined as studies with human subjects\footnote{\url{https://grants.nih.gov/grants/glossary.htm\#ClinicalResearch}}. The presence of the term ``Humans''\footnote{Unique ID D006801; \url{https://meshb.nlm.nih.gov/record/ui?ui=D006801}} in a paper can be used as a heuristic that the paper is about clinical research.

Based on this idea, \citet{Weber-identify-2013} partitioned MEDLINE papers into seven categories, according to whether cell-, animal-, and human-related terms are present in the list of MeSH terms of a paper. The most basic papers are those with only cell and animal terms, papers with only human terms are instead considered as the most clinical, and papers with both are in between. Similarly, \citet{Li-applied-2017} assigned NIH grants to different animal kingdom categories, from virus, prokaryote up to primates and humans, based on the MeSH terms that were inferred from the texts of grant abstracts. They found that basic and applied grants are equal likely in generating papers that are subsequently cited by patents.

\subsection{Citation difference by research level}

Our research question is how the research level of a paper is associated with the number of citations it receives. Despite a few proposals of indicators for research level have been introduced as described above, this question has remained largely unexplored, and existing works seem to generate competing answers. Using visualization techniques, \citet{Eck-citation-2013} showed that clinical words are associated with less citations. \citet{Pasterkamp-lost-2016} looked at journals that belong to various stages of the drug development pipeline (drug target discovery, pre-clinical studies, clinical trials, etc.) and found that for journals with the highest IF at each stage, there is a IF valley---journals that are located between basic and clinical research have the lowest IF. This observation reinforces the ``valley of death'' notion \citep{Butler-valley-2008}. \citet{Opthof-diff-2011} examined papers published in \emph{Circulation} in year 1998 and found that clinical papers seem to have more citations than basic papers. Finally, \citet{Fu-using-2010} used content based features as derived from MeSH terms to predict citations of biomedical papers. The fact that there is a predictive power of MeSH terms suggests that there may be indeed citation difference for papers with different terms.

\section{Level score: an indicator for biomedical research level}

To study how research level is linked to citations, we adopted a newly minted measure called ``level score'' (LS) as the indicator for research level \citep{Ke-identify-2019}. LS is a paper-level indicator that quantifies the extent to which a biomedical paper is oriented towards basic science or clinical research. By construction, a paper with LS closer to -1 is more oriented towards basic science and 1 clinical research. Compared with previous indicators that are nominal values, LS is continuous. The reason for continuity is because discrete value loses its ability in distinguishing papers with \emph{qualitatively} distinct extent of basicness. For example, more than half of MEDLINE papers are associated with only human related MeSH terms, and existing indicators consider them with the same value of basicness. Their actual research levels, however, can be very broad: Phase I clinical trial papers are more basic than Phase IV papers, which are more basic than nursing papers. LS can instead capture this difference.

\begin{figure*}[t!]
\centering
\includegraphics[trim=0mm 0mm 0mm 0mm, width=\textwidth]{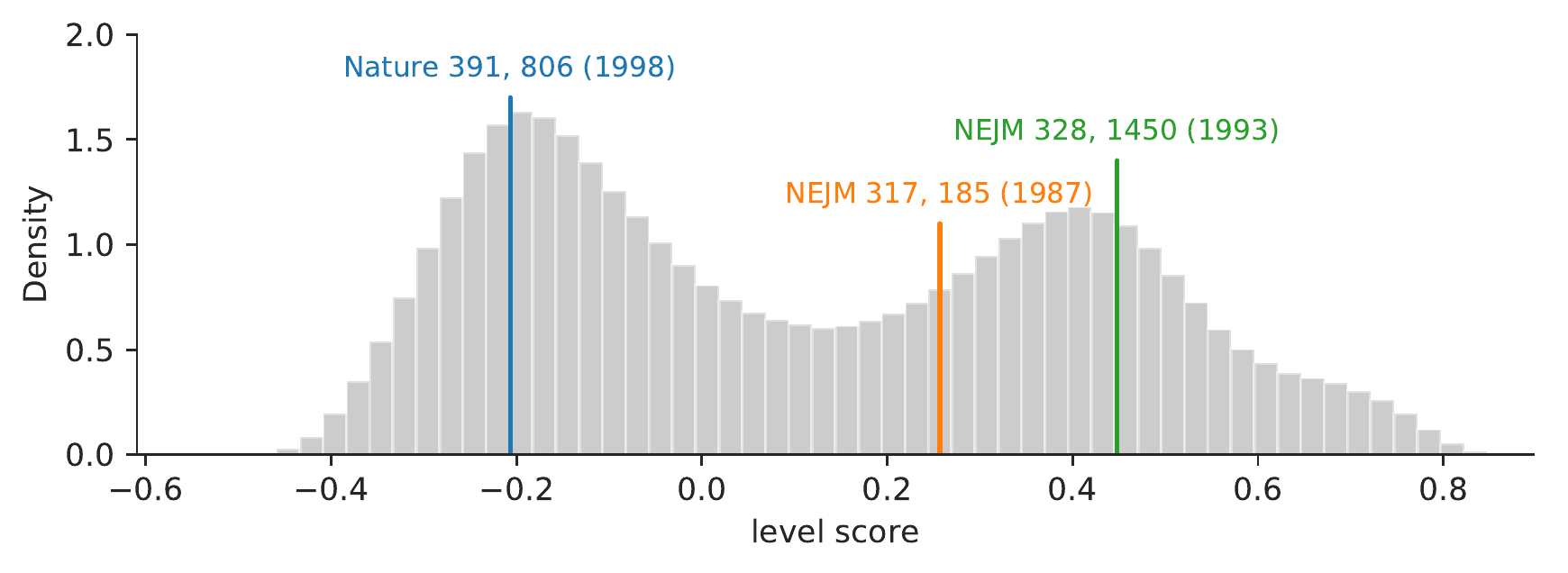}
\caption{Case studies of level scores of individual papers. The histogram shows the distribution of scores for all the 5 million papers included in our sample. Vertical lines mark the scores of three papers, which are, from left to right, (1) \emph{Nature} 391, 806 (1998) \citep{Fire-potent-1998}; (2) \emph{NEJM} 317, 185 (1987) \citep{Fischl-AZT-1987}; and (3) \emph{NEJM} 328, 1450 (1993) \citep{Rimm-vitamin-1993}.}
\label{fig:ls-case}
\end{figure*}

\begin{table*}[t!]
\centering 
\caption{Level score and MeSH terms of three selected papers. Appliedness scores of terms are in parentheses.}
\label{tab:paper-mesh}
\begin{tabular}{l r l}
\hline\hline
\multicolumn{1}{c}{Paper} & \multicolumn{1}{c}{LS} & \multicolumn{1}{c}{MeSH terms} \\ \hline
{\color[HTML]{1f77b4} \href{https://www.ncbi.nlm.nih.gov/pubmed/9486653}{\emph{Nature} 391, 806 (1998)}} & -0.207 & \textbf{Animals} (-0.295); \textbf{Caenorhabditis elegans} (-0.169); \\ 
& & Caenorhabditis elegans Proteins (-0.186); Calmodulin-Binding Proteins (-0.252); \\
& & Gene Expression Regulation (-0.291); Genes, Helminth (-0.167); \\
& & Helminth Proteins (-0.231); Muscle Proteins (-0.223); Phenotype (-0.038); \\
& & RNA, Antisense (-0.307); RNA, Double-Stranded (-0.116) \\ \hline
{\color[HTML]{ff7f0e} \href{https://www.ncbi.nlm.nih.gov/pubmed/3299089}{\emph{NEJM} 317, 185 (1987)}} & 0.257 & AIDS-Related Complex (0.228); Acquired Immunodeficiency Syndrome (0.447); \\
& & Antiviral Agents (-0.098); Clinical Trials as Topic (0.486); \\
& & Double-Blind Method (0.428); Female (--); \textbf{Humans} (0.580); Male (--); \\
& & Opportunistic Infections (0.262); Pneumonia, Pneumocystis (0.225); \\
& & Random Allocation (0.520); Sarcoma, Kaposi (0.253); \\
& & Thymidine (-0.348); Zidovudine (0.098) \\ \hline
{\color[HTML]{2ca02c} \href{https://www.ncbi.nlm.nih.gov/pubmed/8479464}{\emph{NEJM} 328, 1450 (1993)}} & 0.448 & Adult (0.631); Ascorbic Acid (-0.050); Carotenoids (0.078); \\
& & Confidence Intervals (0.748); Coronary Disease (0.450); Diet (0.224); \\
& & Follow-Up Studies (0.631); \textbf{Humans} (0.616); Male (--); Middle Aged (0.578); \\
& & Prospective Studies (0.687); Risk Factors (0.754); Vitamin E (0.030) \\
\hline\hline
\multicolumn{3}{p{\textwidth}}{\footnotesize No appliedness scores are assigned to ``Male'' and ``Female,'' because they were excluded from the calculation, as they do not have tree numbers, which are required to check whether terms are cell-, animal-, or human-related.}\\
\end{tabular}
\end{table*}

Here we briefly describe the main idea behind the calculation of LS and show the values for several exemplar papers, and the details of the calculation are provided in \citet{Ke-identify-2019}. The derivation of LS is still based on the idea that certain MeSH terms indicates the level of experiment and is inspired from recent advances in representation learning in machine learning literature. In particular, it has been shown that vector-representations of words that are learned from large-scale text corpora capture semantic relationships between words, as exemplified by the equation $v(\text{king}) - v(\text{man}) + v(\text{woman}) \approx v(\text{queen})$, where $v(\cdot)$ denotes the vector of a word \citep{Mikolov-linguistic-2013, Mikolov-efficient-2013}. From these vectors, one can induce the gender dimension by vector difference of pairs of words; a simple arithmetic operation of the above equation gives us $v(\text{king}) - v(\text{queen}) \approx v(\text{man}) - v(\text{woman})$, where both $v(\text{king}) - v(\text{queen})$ and $v(\text{man}) - v(\text{woman})$ are instances of the gender axis. Averaging over all these pairs of words (e.g., male and female) yields the gender dimension. One can then project a word onto the gender dimension to study the extent of its similarity to male and female. For example, studies have shown that---based on vectors learned from text corpora---computer programmer is more to male than female, and further found that this type of analogy reflects societal bias, which can be quantified through word vectors \citep{Garg-word-2018}.

The calculation of LS resembles the process described above. First, popular methods are used to learn vector-representations of MeSH terms\footnote{MeSH terms are organized into a hierarchy with 16 branches, and most terms are associated with one or more tree numbers, indicating their locations in the tree structure. Only terms that have tree numbers and are located in branches A--E, G, M, and N are considered.}. Then, in the same way for identifying the gender axis, the translation axis (TA)---a vector pointing from basic science MeSH terms to clinical research terms---is identified in the vector space, where cell- and animal-related terms are used as basic science terms and human-related terms as clinical. Next, each MeSH term are projected onto TA to obtain its ``appliedness'' score. The LS of a paper is the average appliedness score of its MeSH terms, indicating averaged position of its MeSH terms on TA.

Note that when calculating LS of a focal paper, only MeSH terms that appeared in papers published in the last 5 years were considered. This is based on the consideration that biomedical research has been evolving. For example, \citet{Ke-identify-2019} showed some MeSH terms whose appliedness scores have exhibited noticeable changes during the studied period. One may, therefore, concern if the scores have also changed for some very general terms such as ``Humans.'' Figure~\ref{fig:mesh-score} shows that this is not the case; its appliedness score remains relatively stable over time, exhibiting a similar trend to the average of all terms.

To demonstrate the validity of LS, let us present some case studies of individual papers. In Figure~\ref{fig:ls-case}, we first show the histogram of LS for all the 5 million papers in our sample (see \S~\ref{subsec:sample} for sample selection), serving as the background distribution against which we compare with several exemplar papers. The histogram exhibits bimodality, with the two modes corresponding to basic science and clinical medicine. Bimodality also indicates relatively less amount of published translational science papers compared to either basic or clinical papers, reflecting the gap between basic and clinical science.

On top of the histogram, we highlight LS of the following papers. The first one [{\color[HTML]{1f77b4} \href{https://www.ncbi.nlm.nih.gov/pubmed/9486653}{\emph{Nature} 391, 806 (1998)}}] discovered the RNA interference phenomenon and wined the 2006 Nobel Prize in Physiology or Medicine \citep{Fire-potent-1998}. The first row in Table~\ref{tab:paper-mesh} lists the 11 MeSH terms associated to the paper and their appliedness scores in parentheses. Two terms, namely ``Animals'' and ``Caenorhabditis elegans,'' describe that the research was conducted on the \emph{Caenorhabditis elegans} model organism\footnote{``Caenorhabditis elegans'' is a descendant node of ``Animals'' in the MeSH tree.}, and hence the paper is considered as basic research using previous indicators. This is consistent with its LS, as suggested from Figure~\ref{fig:ls-case} showing that the LS of the paper is located around the peak of the histogram that corresponds to basic science, therefore indicating that it is a typical basic science paper.

The second paper [{\color[HTML]{ff7f0e} \href{https://www.ncbi.nlm.nih.gov/pubmed/3299089}{\emph{NEJM} 317, 185 (1987)}}] is a highly-cited, clinical trial study that tested the effectiveness of Zidovudine, a medication for HIV infection \citep{Fischl-AZT-1987}. The second row in Table~\ref{tab:paper-mesh} enumerates its terms, and the fact that ``Humans'' is one of them means that the paper is categorized as clinical medicine by existing indicators. The LS of the paper also indicates that it is a clinical paper, as seen from Figure~\ref{fig:ls-case}. Yet, LS can further capture that compared with typical clinical papers, it is more orientated toward basic science. This is because the paper is about drug development, which follows the pipeline that active drug ingredients first need to test on animals before entering the clinical trial stage.

The third paper [{\color[HTML]{2ca02c} \href{https://www.ncbi.nlm.nih.gov/pubmed/8479464}{\emph{NEJM} 328, 1450 (1993)}}] is a highly-cited, prospective study on the association between dietary antioxidants consumption and the risk of primary coronary artery disease in men \citep{Rimm-vitamin-1993}. Its MeSH terms contain ``Humans,'' it is therefore classified as clinical research by existing indicators, as well as by its LS. However, this paper is a typical epidemiology study, without testing new drugs, thus is actually considered as less basic than the second paper. This can be captured by LS, as can be seen from Figure~\ref{fig:ls-case} that its LS is larger than that of the second paper.

Going beyond individual examples, \citet{Ke-identify-2019} presented systematic validations of LS. For example, papers that are labeled as clinical trials are more oriented towards clinical research, with level scores getting progressively larger for papers of later stages clinical trial. Aggregating LS onto the field level, it was shown that cell biology and biochemistry are the most basic fields, whereas nursing the most clinical.

\section{Data and methods} \label{sec:data}

\subsection{Sample selection} \label{subsec:sample}

To address our research question, we analyze papers indexed in MEDLINE, a publicly available database curated by the NLM for biomedical research literature. Our unit of analysis is a scientific paper. The papers in our sample were published between 1980 and 2002, so that every paper has at least 15 years to garner citations. We use Web of Science (WoS) data to obtain additional information about papers. In particular, we limit our attention to papers that are designated as articles in WoS and further get their field labels, as specified as WoS subject category, and their total citation counts, regardless of the fields where citing papers are from. Our sample contains over 5 million papers. Table~\ref{tab:field} presents the top 30 largest fields based on the number of papers in the sample. These fields accounts for 78.2\% of all papers in the sample.

\subsection{Dependent variables}

The number of forward citations a paper receives has been widely used as an indicator for its scientific impact. We introduce two groups of dependent variables. The first group is the total number of forward citations that a focal paper has accrued within the 3-, 10-, and 15-year windows after its publication, denoted as $c3$, $c10$, and $c15$, respectively. We consider different lengths of citation-windows to account for the paper- and time-dependent tendency to acquire citations.

Since citations are heterogeneously distributed, spanning orders of magnitude (Table~\ref{tab:var}), to capture the right tail of citation distribution, our second group of dependent variables is binary variables that indicates whether a focal paper is a hit one, defined as its total citation count is ranked into the top 1\% among all papers in the same field and the same publication year. We use the same three window lengths, and the three dependent variables are similarly denoted as $h3$, $h10$, and $h15$.

\subsection{Independent variables}

Our main independent variable of interest is the level score of a focal paper, denoted as $ls$.

In light of prior research, we consider the following control variables. The first is the number of authors, as previous studies have repeatedly demonstrated the increasing prevalence of collaboration in knowledge production \citep{Wuchty-team-2007} and the positive association between collaboration and citations \citep{Bornmann-what-2012, Peng-where-2012, Didegah-determinant-2013, Fanelli-positive-2013}. The second control variable is whether a focal paper involves international collaboration, which has been shown to have a positive link to citations \citep{Sin-intl-2011}. The third is the number of MeSH terms, as it has a positive effect on citations \citep{Chai-breakthrough-2019}.

Table~\ref{tab:var} reports the summary statistics of all the variables introduced. Figure~\ref{fig:var-dist} shows the distributions of non-indicator variables except LS, which has been shown in Figure~\ref{fig:ls-case}.

Aside from the above control variables, we include the field and publication year fixed effects. Therefore, our estimations of the effect of research level on citations are within the same field but not across fields. Publication year fixed effect is included to address the concern that ignoring year would make it an omitted variable; it controls for characteristics that are fixed within a year but vary across time, such as the number of publications and the amount of funding.

\subsection{Model specification}

We use generalized negative binomial (GNB) regression, as provided by the \texttt{gnbreg} function in \textsc{STATA}, to model the association between research level and citation count. We choose GNB because (1) it can model over-dispersed count variable, which is the case for citation count, and (2) it can model not only the mean but also the dispersion parameter. The model specification is as follows:
\begin{equation}
c_i = \beta_0 + \beta_1 \cdot ls_i + \gamma \cdot \text{controls}_i + \delta_i + \eta_i + \varepsilon_i \, ,
\end{equation}
where $\beta_1$ is the coefficient of interest for the research level independent variable, $\delta_i$ is a dummy variable for WoS subject category (field fixed effect), $\eta_i$ is a dummy for publication year, and $\varepsilon_i$ is the noise term.

We employ logistic regression model to estimate the effect of level score on being a hit paper, controlling for other confounding variables. The model specification is similar to the GNB one.

\section{Results} \label{sec:res}

\subsection{Clinical papers are undercited}

Table~\ref{tab:c-stat} groups papers into 10 categories base on their LS and show mean and standard deviation summary statistics of citations for papers in each group. We observe that average number of citations decreases as we move along the spectrum from basic to clinical research.

We further test whether the observation showed in Table~\ref{tab:c-stat} holds after controlling for other confounding factors. We run negative binomial regressions to estimate 3-, 10-, and 15-year citations of papers and report the full regression results in Tables~\ref{tab:c3m}--\ref{tab:c15m}. These results fully support Table~\ref{tab:c-stat}, reassuring a negative association between research level and citation counts.

\begin{table}[t!]
\centering 
\caption{Mean and standard deviation of citations for papers grouped by their level scores.}
\label{tab:c-stat}
\begin{tabular}{lrrrrrr}
\hline\hline
{}                 & \multicolumn{2}{c}{$c3$} & \multicolumn{2}{c}{$c10$} & \multicolumn{2}{c}{$c15$} \\
Level score range  & Mean & Std. Dev. & Mean & Std. Dev. & Mean & Std. Dev. \\
\hline
$< -0.308$ & 13.691 & 24.619 & 34.425 & 66.751 & 41.522 & 84.190 \\
$< -0.007$ & 10.766 & 20.601 & 29.606 & 69.053 & 37.723 & 98.234 \\
$< 0.295$ &  6.706 & 13.976 & 20.015 & 48.280 & 26.372 & 80.252 \\
$< 0.596$ &  5.049 & 11.702 & 16.736 & 39.610 & 22.859 & 58.260 \\
$\geq 0.596$ &  4.445 & 10.034 & 17.241 & 39.490 & 25.199 & 67.556 \\
\hline\hline
\end{tabular}
\end{table}

Let us first describe the results for 3-year citations (Table~\ref{tab:c3m}). Model 1 in Table~\ref{tab:c3m} indicates that after controling only for the publication year fixed effect, a one-unit increase in level score is linked to $73\%$ ($e^{-1.309} - 1$) decrease of $c3$. Model 2 in Table~\ref{tab:c3m} suggests that after controlling for the field fixed effect, for each one-unit increase in level score, the expected citation count in 3 years decreases by $39\%$. If we additionally include the publication year fixed effect, the effect size is more pronounced---$49\%$ decrease, as shown in Model 3. In Model 4, we further consider all independent control variables and find that one-unit increase in LS translates to $41\%$ significant decrease of 3-year citations. As side results, Model 4 reinforces several previous studies; that is, both the number of authors and international collaboration are positively linked to the number of citations.

Next, we explore how the effect of research level on citations may have changed over time. In doing so, we divide the full sample into three disjoint subsamples, each corresponding to the decade of publication year. Models 5--7 in Table~\ref{tab:c3m} display the modeling results. We find that the effect size decreases over time; for papers published in 1980s, one-unit increase in level score is associated with a sizable ($52\%$) decrease of 3-year citations, and such negative effect had weakened to $39\%$ in 1990s and $34\%$ in 2000--2002. Understanding factors behind this shrinking effect is beyond the scope of this work, but serves as an important future study. We nevertheless conjecture that an account for this is the growing importance of clinical medicine.

\begin{figure}[t!]
\centering
\includegraphics[trim=0mm 0mm 0mm 0mm, width=0.6\columnwidth]{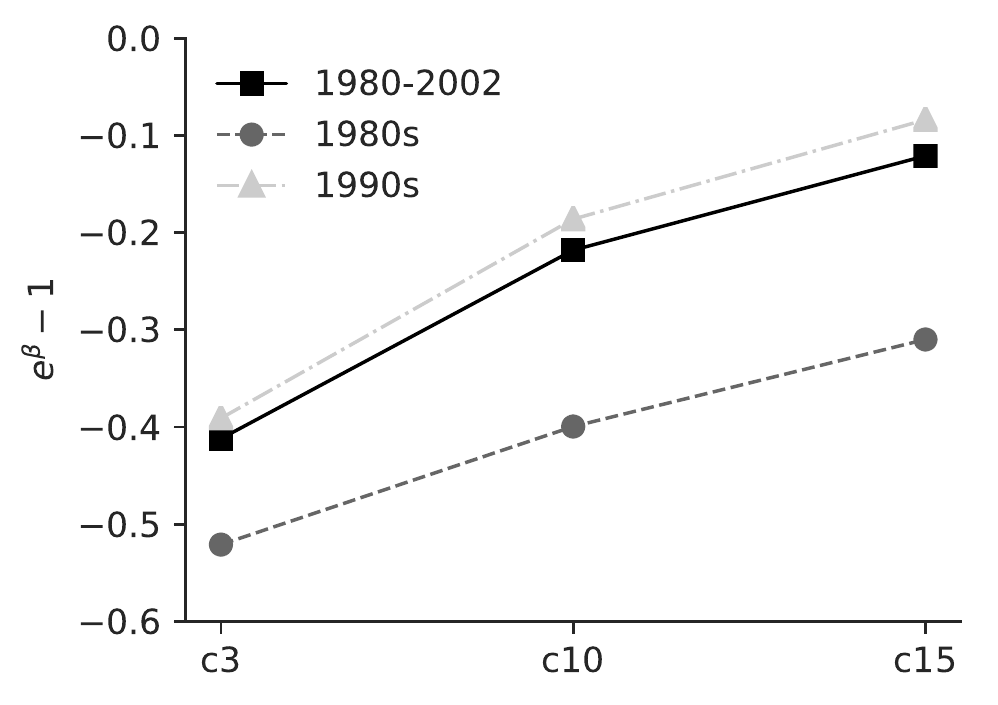}
\caption{Effect sizes of level score on citations. In all cases, we use negative binomial regression considering field and publication year fixed effects and other control variables.}
\label{fig:beta-ls}
\end{figure}

The results presented so far are based on 3-year citations. To understand how the effect of research level on citations may vary for different lengths of citation windows, we separately model citations in 10 and 15 years, which serve as the proxies for medium- and long-term scientific impact, whereas 3-year citation is for short-term impact. In both cases, we consider the full sample as well as  subsamples of papers grouped by decade. Tables~\ref{tab:c10m} and \ref{tab:c15m} respectively present the results for 10- and 15-year citations. To facilitate comparisons, we summarize all the modeling results in Figure~\ref{fig:beta-ls}, where we plot the effect sizes ($e^{\beta_1} - 1$) of LS on citations.

Figure~\ref{fig:beta-ls} first indicates that LS consistently has a negative effect on citations. This means that papers that are more orientated towards clinical medicine tend to acquire fewer citations. The extent, however, varies across time and citation-window length. Looking at the full sample, Figure~\ref{fig:beta-ls} (see also Models 4 in Tables~\ref{tab:c3m}--\ref{tab:c15m}) suggests a stronger effect for shorter citation window. In particular, one-unit increase in LS is associated with $41\%$ decrease of 3-year citations ($c3$), $22\%$ decrease of $c10$ and $12\%$ decrease of $c15$. Turning to the decade subsamples, we observe a similar weakening influence of research level on citations as we increase window length. For papers published in 1980s, the size of the negative effect decreases from $52\%$ for $c3$ to $40\%$ for $c10$ to $31\%$ for $c15$ (Models 5 in Tables~\ref{tab:c3m}--\ref{tab:c15m}). For papers in 1990s, the size shrinks from $39\%$ for $c3$ to $19\%$ for $c10$ to $8\%$ for $c15$ (Models 6 in Tables~\ref{tab:c3m}--\ref{tab:c15m}). These results signify that life-cycles of papers' citations may differ by their research levels: Shortly after publication, citations of clinical orientated papers may be less saturated, we thus observe a strong effect of research level on citations; as we increase the citation window, their citations become more saturated, and the effect correspondingly becomes weaker.

Aside from modeling mean citations, we also examine the dispersion of citations. Tables~\ref{tab:c3v}--\ref{tab:c15v}, which provide estimates from GNB models, demonstrate a consistently positive association between LS and dispersion of citations, regardless of window length and publication time. This implies that clinical oriented papers have higher dispersion in citations.

\subsection{Probability of hit papers}

\begin{figure}[t!]
\centering
\includegraphics[trim=0mm 0mm 0mm 0mm, width=0.6\columnwidth]{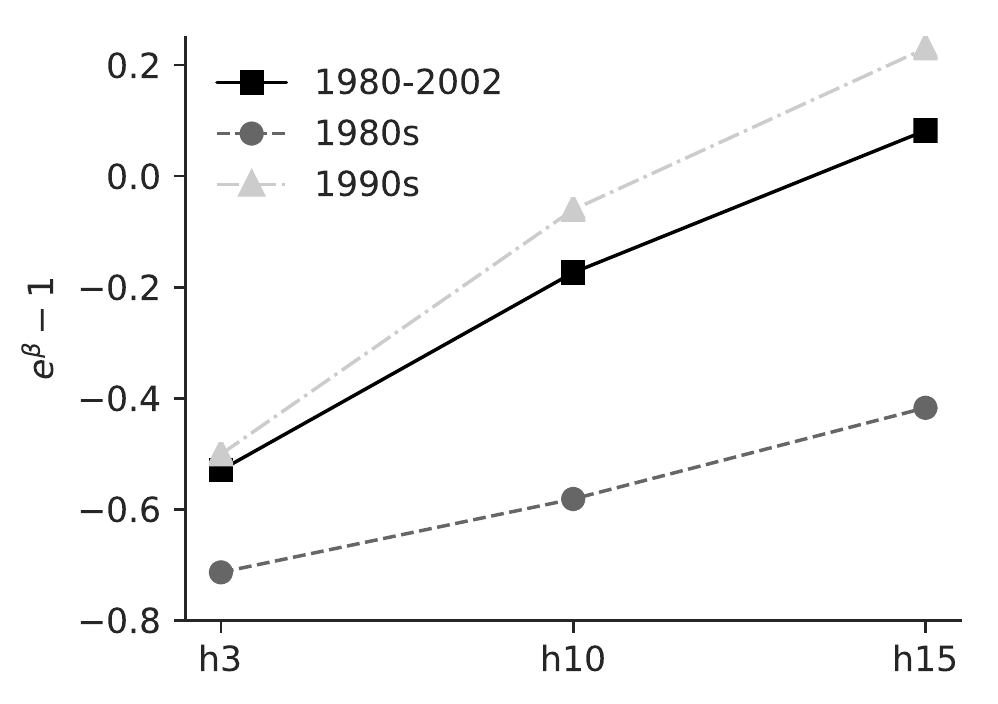}
\caption{Effect sizes of level score on being ranked into the top 1\% most cited papers in the same field and year. In all cases, we use logistic regression considering field and publication year fixed effects and other control variables.}
\label{fig:beta-ls-h}
\end{figure}

As citation count exhibits over-dispersion, we investigate the right tail of the distributions and model the probability of breakthrough papers. We again use 3-, 10-, and 15-year citations as the short-, medium-, and long-term scientific impact, and utilize logistic regression to model whether a paper is among the top 1\% most cited ones clustered by field and publication year. Similar to the regression case, we summarize all the effect size results of LS in Figure~\ref{fig:beta-ls-h}, and model estimates are instead detailed in Tables~\ref{tab:h3}--\ref{tab:h15}.

Focusing on the full sample, Figure~\ref{fig:beta-ls-h} (see also Model 4 in Table~\ref{tab:h3}) shows that after controlling for field and year fixed effects and all other confounders, a one-unit increase of LS is linked to a $53\%$ ($e^{-0.754}$ - 1) significant decrease in the odds of being made in the top 1\% most cited papers based on 3-year citations. For 10-year citations, LS is still negatively correlated with being a ``hit'' paper, but the effect size is much smaller than the 3-year citation case, reaching to only $17\%$ (Model 4 in Table~\ref{tab:h10}). As we continue to expand the length of citation-window to 15 years, we observe that there is actually a $8\%$ increase in the odds of becoming hit papers (Model 4 in Table~\ref{tab:h15}).

We explore time trends of the effect of research level on being a hit paper. For papers in 1980s, we find a negative yet weakening effect. Each one-unit increase of LS translates to $71\%$, $58\%$, and $42\%$ significant decrease in the odds of being recognized as breakthroughs after 3, 10, and 15 years of publication, respectively. For papers in the 1990s, we see a negative-to-positive transition of the effect, similar to the full sample case.

\section{Concluding discussion} \label{sec:dis}

Biomedical research has two broad goals: advancing the understandings of living organisms through basic science and improving human health through clinical research. The purpose of this work was to understand whether there is any systematic differences in the amount of citations received for the two types of research. This is important, because citation-based indicators for scientific impact have been increasingly used for research evaluation. We have used a paper-level indicator that quantifies the extent of basicness and studied its association with short-, medium-, and long-term citations for a large-scale corpus of more than 5 million papers published in 1980-2002. Our analysis has pointed out the disadvantage of clinical orientated research in accruing citations; clinical papers tend to have fewer citations than basic papers and are less likely to become hit works. We have further found that such difference is more pronounced for short-term citations.

These results may speak to the policy literature. Bio-researchers appear to be facing a dilemma posing by current science policy. On one hand, a proven record of individual excellence in publications and citations is underscored in funding and tenure decisions. On the other, the public, government, and funding bodies have been increasingly calling for researchers to demonstrate the direct impact of their research on the improvement of human health and the societal impact in general. The demonstration of clinical relevance requires the translation of basic scientific discoveries into clinical applications, from which patients can benefit. This process, however, may take decades with frequent failures \citep{Contopoulos-life-2008}, which results in less number of publications and less amount of citations accrued, which in turn makes translational researchers less competitive in evaluation. This may provide disincentive to researchers to work on the translation of basic science, resulting in a gap between basic science and clinical research that is coined as ``valley of death'' \citep{Butler-valley-2008}. Given the current citation under-appreciation of clinical research, funding bodies may seek to have comprehensive evaluation of output beyond scholarly publications and short-term based citations. Examples can be looking at citations in other types of documents such as clinical guidance \citep{Grant-evaluating-2000} and drug development documents \citep{Williams-from-2015}.

Our work also adds to the bibliometrics literature that has studied potential biases in citation data. One of the most well-known biases is the field bias \citep{Bornmann-what-2008}; that is, papers in certain fields on average have more citations than other fields. Reasons for such difference may include varying number of papers and number of references in a paper. Many methods have been proposed in attempting to alleviate this problem \citep{Radicchi-universality-2008}. The key idea behind these methods is to compare a paper's raw number of citations with citations of papers in the same field, resulting in field-normalized indicators. This procedure considers cross-field citation difference only, thus fails to account for within-filed case, which, as we have shown, happens for papers that are in the same field but belong to different types of researches. Field-normalized indicators, therefore, are not able to correct research level bias. Proposing citation indicators that can mitigate such bias remains to be an important future work.

\section*{Acknowledgments}

Computing resources and the \textsc{Stata} software provided by Northeastern University are gratefully acknowledged.

\appendix

\setcounter{figure}{0}
\makeatletter 
\renewcommand{\thefigure}{A.\@arabic\c@figure}
\makeatother

\setcounter{table}{0}
\makeatletter 
\renewcommand{\thetable}{A.\@arabic\c@table}
\makeatother

\section{}

\begin{figure}[ht!]
\centering
\includegraphics[trim=0mm 2mm 0mm 0mm, width=0.8\textwidth]{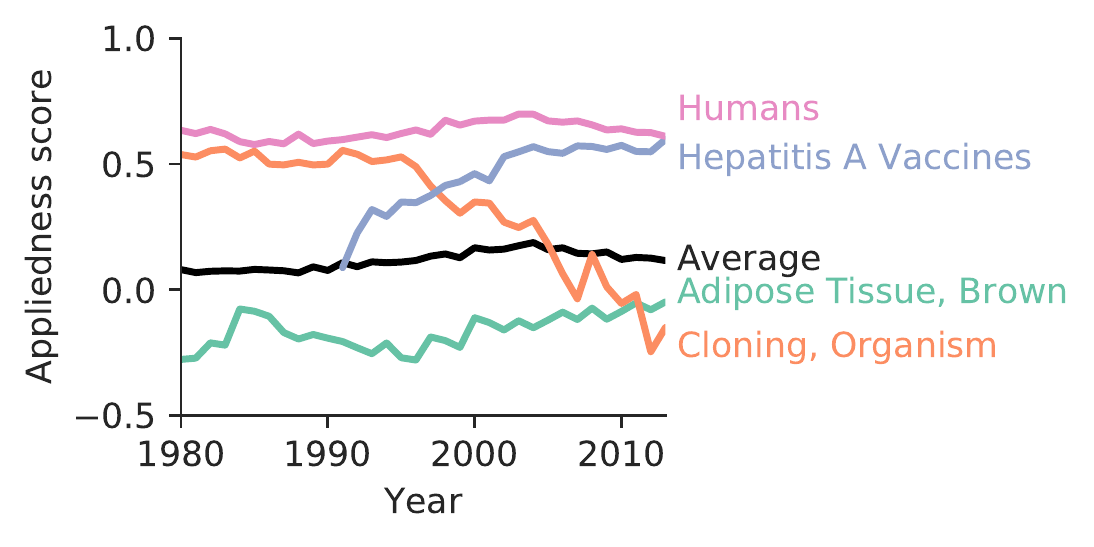}
\caption{Appliedness scores of selected MeSH terms over time.}
\label{fig:mesh-score}
\end{figure}

\begin{figure*}[ht!]
\centering
\includegraphics[trim=0mm 2mm 0mm 0mm, width=\textwidth]{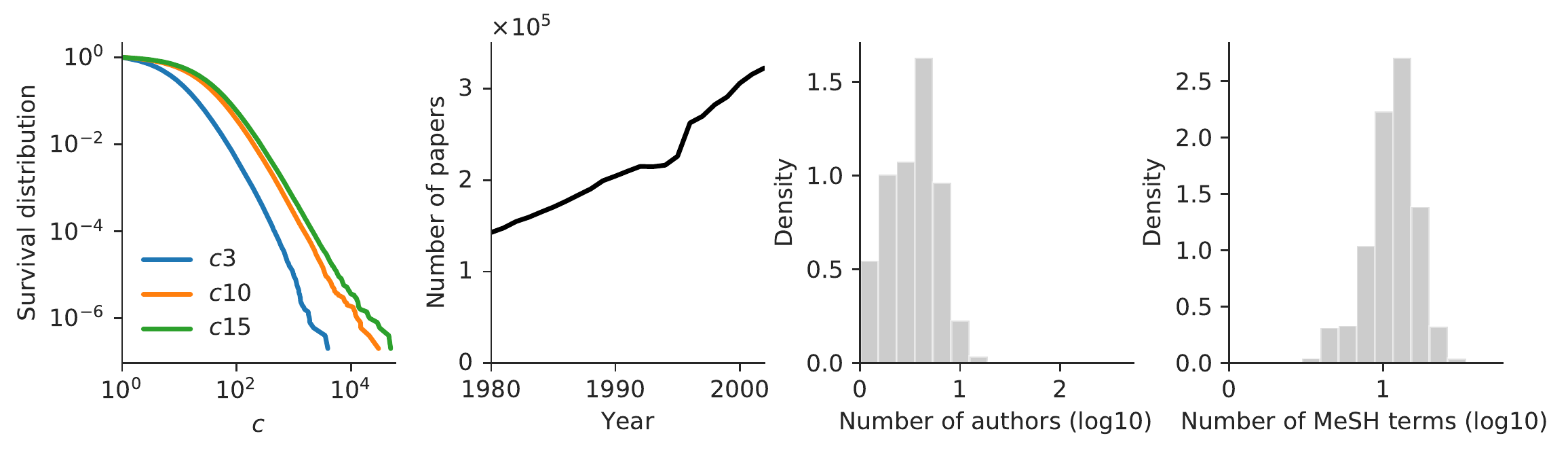}
\caption{Distributions of variables.}
\label{fig:var-dist}
\end{figure*}

\begin{table}[t!]
\centering
\caption{The number and percentage of papers in the top 30 largest fields in our corpus.}
\label{tab:field}
\begin{tabular}{lrr}
\hline\hline
                                         Field &  Papers &  Percentage \\
\hline
             Biochemistry \& Molecular Biology &  524373 &       10.44 \\
                 Medicine, General \& Internal &  263400 &        5.24 \\
                      Pharmacology \& Pharmacy &  228988 &        4.56 \\
                                 Neurosciences &  222744 &        4.43 \\
                                       Surgery &  221762 &        4.41 \\
                                    Immunology &  199624 &        3.97 \\
                                      Oncology &  198047 &        3.94 \\
             Cardiac \& Cardiovascular Systems &  146685 &        2.92 \\
                   Endocrinology \& Metabolism &  134664 &        2.68 \\
Radiology, Nuclear Medicine \& Medical Imaging &  121995 &        2.43 \\
                                  Cell Biology &  117688 &        2.34 \\
  Public, Environmental \& Occupational Health &  108608 &        2.16 \\
                            Clinical Neurology &  101155 &        2.01 \\
                                  Microbiology &  100888 &        2.01 \\
                          Genetics \& Heredity &   97359 &        1.94 \\
                    Multidisciplinary Sciences &   93317 &        1.86 \\
                                    Physiology &   92344 &        1.84 \\
                         Urology \& Nephrology &   85991 &        1.71 \\
                      Obstetrics \& Gynecology &   82911 &        1.65 \\
                                    Pediatrics &   81059 &        1.61 \\
                           Veterinary Sciences &   80957 &        1.61 \\
                                 Ophthalmology &   79952 &        1.59 \\
                Gastroenterology \& Hepatology &   79522 &        1.58 \\
                                    Hematology &   75930 &        1.51 \\
                                     Pathology &   72687 &        1.45 \\
                                    Psychiatry &   72197 &        1.44 \\
           Dentistry, Oral Surgery \& Medicine &   69106 &        1.38 \\
                                   Dermatology &   61867 &        1.23 \\
            Medicine, Research \& Experimental &   60229 &        1.20 \\
                                      Virology &   53014 &        1.06 \\
\hline\hline
\end{tabular}
\end{table}

\begin{table*}[t]
\centering 
\caption{Summary statistics for all the variables used in our analysis.}
\label{tab:var}
\begin{tabular}{l r r r r r}
\hline\hline
Variable & \multicolumn{1}{c}{Mean} & \multicolumn{1}{c}{Std. Dev.} & \multicolumn{1}{c}{Min} & \multicolumn{1}{c}{Max} & \multicolumn{1}{c}{N} \\ \hline
c3 & 8.033 & 17.068 & 0 & 3933 & $5\,026\,235$ \\
c10 & 23.387 & 56.281 & 0 & 30327 & $5\,026\,235$ \\
c15 & 30.504 & 82.453 & 0 & 49315 & $5\,026\,235$ \\
h3 & 0.0104 & 0.101 & 0 & 1 & $4\,985\,532$ \\
h10 & 0.010 & 0.100 & 0 & 1 & $4\,985\,532$ \\
h15 & 0.010 & 0.100 & 0 & 1 & $4\,985\,532$ \\
year & 1992.631 & 6.563 & 1980 & 2002 & $5\,026\,235$ \\
ls & 0.114 & 0.310 & -0.609 & 0.897 & $5\,026\,235$ \\
authornum & 4.082 & 2.690 & 1 & 546 & $5\,026\,235$ \\
authorintl & 0.109 & 0.312 & 0 & 1 & $4\,112\,865$ \\
meshnum & 12.19 & 4.447 & 1 & 60 & $5\,026\,235$ \\
\hline\hline
\end{tabular}
\end{table*}

\begin{table*}[t]
\centering
\caption{Negative binomial regression modeling of 3-year citations.}
\label{tab:c3m}
\begin{tabular}{l*{7}{c}}
\hline\hline
              & (1) & (2) & (3) & (4) & (5) & (6) & (7) \\
c3            & 1980--2002 & 1980--2002 & 1980--2002 & 1980--2002 & 1980--1989 & 1990--1999 & 2000--2002 \\
\hline
ls            & -1.309\sym{***} & -0.496\sym{***} & -0.678\sym{***} & -0.531\sym{***} & -0.736\sym{***} & -0.496\sym{***} & -0.409\sym{***} \\
              & (0.00191)       & (0.00274)       & (0.00280)       & (0.00294)       & (0.00587)       & (0.00412)       & (0.00599)       \\
[1em]
ln\_authornum &                 &                 &                 & 0.310\sym{***}  & 0.365\sym{***}  & 0.295\sym{***}  & 0.264\sym{***}  \\
              &                 &                 &                 & (0.00101)       & (0.00194)       & (0.00145)       & (0.00206)       \\
[1em]
meshnum       &                 &                 &                 & 0.0427\sym{***} & 0.0604\sym{***} & 0.0432\sym{***} & 0.0358\sym{***} \\
              &                 &                 &                 & (0.000134)      & (0.000333)      & (0.000188)      & (0.000234)      \\
[1em]
authorintl    &                 &                 &                 & 0.146\sym{***}  & 0.105\sym{***}  & 0.160\sym{***}  & 0.161\sym{***}  \\
              &                 &                 &                 & (0.00174)       & (0.00460)       & (0.00242)       & (0.00297)       \\
[1em]
Constant      & 1.810\sym{***}  & 1.643\sym{***}  & 1.567\sym{***}  & 0.899\sym{***}  & 0.994\sym{***}  & 0.697\sym{***}  & 1.014\sym{***}  \\
              & (0.00363)       & (0.0532)        & (0.0535)        & (0.0604)        & (0.109)         & (0.0918)        & (0.108)         \\
\hline
Field fe      &                 & $\checkmark$    & $\checkmark$    & $\checkmark$    & $\checkmark$    & $\checkmark$    & $\checkmark$    \\
Year fe       & $\checkmark$    &                 & $\checkmark$    & $\checkmark$    & $\checkmark$    & $\checkmark$    & $\checkmark$    \\
Observations  & 5024732         & 5024732         & 5024732         & 4111637         & 1172896         & 1997524         & 941217          \\
\hline\hline
\multicolumn{7}{l}{\footnotesize Standard errors in parentheses}\\
\multicolumn{7}{l}{\footnotesize \sym{*} \(p<0.05\), \sym{**} \(p<0.01\), \sym{***} \(p<0.001\)}\\
\end{tabular}
\end{table*}

\begin{table*}[t]
\centering
\caption{Negative binomial regression modeling of 10-year citations.}
\label{tab:c10m}
\begin{tabular}{l*{7}{c}}
\hline\hline
              & (1) & (2) & (3) & (4) & (5) & (6) & (7) \\
c10           & 1980--2002 & 1980--2002 & 1980--2002 & 1980--2002 & 1980--1989 & 1990--1999 & 2000--2002 \\
\hline
ls            & -0.963\sym{***} & -0.185\sym{***} & -0.397\sym{***} & -0.246\sym{***} & -0.510\sym{***} & -0.206\sym{***} & -0.0884\sym{***} \\
              & (0.00181)       & (0.00265)       & (0.00270)       & (0.00283)       & (0.00574)       & (0.00392)       & (0.00579)        \\
[1em]
ln\_authornum &                 &                 &                 & 0.293\sym{***}  & 0.348\sym{***}  & 0.280\sym{***}  & 0.246\sym{***}   \\
              &                 &                 &                 & (0.000980)      & (0.00191)       & (0.00138)       & (0.00199)        \\
[1em]
meshnum       &                 &                 &                 & 0.0429\sym{***} & 0.0632\sym{***} & 0.0434\sym{***} & 0.0353\sym{***}  \\
              &                 &                 &                 & (0.000130)      & (0.000327)      & 0.000181)       & (0.000229)       \\
[1em]
authorintl    &                 &                 &                 & 0.144\sym{***}  & 0.0991\sym{***} & 0.155\sym{***}  & 0.164\sym{***}   \\
              &                 &                 &                 & (0.00171)       & (0.00461)       & (0.00236)       & (0.00291)        \\
[1em]
Constant      & 2.832\sym{***}  & 2.934\sym{***}  & 2.836\sym{***}  & 2.177\sym{***}  & 2.155\sym{***}  & 2.094\sym{***}  & 2.588\sym{***}   \\
              & (0.00360)       & (0.0519)        & (0.0517)        & (0.0576)        & (0.105)         & (0.0852)        & (0.107)          \\
\hline
Field fe      &                 & $\checkmark$    & $\checkmark$    & $\checkmark$    & $\checkmark$    & $\checkmark$    & $\checkmark$     \\
Year fe       & $\checkmark$    &                 & $\checkmark$    & $\checkmark$    & $\checkmark$    & $\checkmark$    & $\checkmark$     \\
Observations  & 5024732         & 5024732         & 5024732         & 4111637         & 1172896         & 1997524         & 941217           \\
\hline\hline
\multicolumn{7}{l}{\footnotesize Standard errors in parentheses} \\
\multicolumn{7}{l}{\footnotesize \sym{*} \(p<0.05\), \sym{**} \(p<0.01\), \sym{***} \(p<0.001\)} \\
\end{tabular}
\end{table*}

\begin{table*}[t]
\centering
\caption{Negative binomial regression modeling of 15-year citations.}
\label{tab:c15m}
\begin{tabular}{l*{7}{c}}
\hline\hline
              & (1) & (2) & (3) & (4) & (5) & (6) & (7) \\
c15           & 1980--2002 & 1980--2002 & 1980--2002 & 1980--2002 & 1980--1989 & 1990--1999 & 2000--2002 \\
\hline
ls            & -0.835\sym{***} & -0.0394\sym{***} & -0.277\sym{***} & -0.129\sym{***} & -0.371\sym{***} & -0.0876\sym{***} & 0.0117\sym{*}   \\
              & (0.00182)       & (0.00268)        & (0.00272)       & (0.00286)       & (0.00579)       & (0.00396)        & (0.00590)       \\
[1em]
ln\_authornum &                 &                  &                 & 0.281\sym{***}  & 0.335\sym{***}  & 0.270\sym{***}   & 0.233\sym{***}  \\
              &                 &                  &                 & (0.000994)      & (0.00193)       & (0.00140)        & (0.00204)       \\
[1em]
meshnum       &                 &                  &                 & 0.0420\sym{***} & 0.0626\sym{***} & 0.0426\sym{***}  & 0.0341\sym{***} \\
              &                 &                  &                 & (0.000132)      & (0.000330)      & (0.000183)       & (0.000234)      \\
[1em]
authorintl    &                 &                  &                 & 0.148\sym{***}  & 0.102\sym{***}  & 0.159\sym{***}   & 0.169\sym{***}  \\
              &                 &                  &                 & (0.00174)       & (0.00466)       & (0.00240)        & (0.00298)       \\
[1em]
Constant      & 3.059\sym{***}  & 3.263\sym{***}   & 3.151\sym{***}  & 2.501\sym{***}  & 2.402\sym{***}  & 2.451\sym{***}   & 3.081\sym{***}  \\
              & (0.00367)       & (0.0528)         & (0.0524)        & (0.0583)        & (0.106)         & (0.0850)         & (0.114)         \\
\hline
Field fe      &                 & $\checkmark$     & $\checkmark$    & $\checkmark$    & $\checkmark$    & $\checkmark$     & $\checkmark$    \\
Year fe       & $\checkmark$    &                  & $\checkmark$    & $\checkmark$    & $\checkmark$    & $\checkmark$     & $\checkmark$    \\
Observations  & 5024732         & 5024732          & 5024732         & 4111637         & 1172896         & 1997524          & 941217          \\
\hline\hline
\multicolumn{7}{l}{\footnotesize Standard errors in parentheses}\\
\multicolumn{7}{l}{\footnotesize \sym{*} \(p<0.05\), \sym{**} \(p<0.01\), \sym{***} \(p<0.001\)}\\
\end{tabular}
\end{table*}

\begin{table*}[t]
\centering
\caption{Negative binomial regression modeling of dispersion of 3-year citations.}
\label{tab:c3v}
\begin{tabular}{l*{7}{c}}
\hline\hline
              & (1) & (2) & (3) & (4) & (5) & (6) & (7) \\
lnalpha       & 1980--2002 & 1980--2002 & 1980--2002 & 1980--2002 & 1980--1989 & 1990--1999 & 2000-2002 \\
\hline
ls            & 0.498\sym{***} & 0.406\sym{***} & 0.499\sym{***} & 0.344\sym{***}   & 0.397\sym{***}   & 0.308\sym{***}   & 0.363\sym{***}   \\
              & (0.00229)      & (0.00352)      & (0.00365)      & (0.00407)        & (0.00805)        & (0.00570)        & (0.00865)        \\
[1em]
ln\_authornum &                &                &                & -0.0863\sym{***} & -0.133\sym{***}  & -0.0636\sym{***} & -0.0651\sym{***} \\
              &                &                &                & (0.00140)        & (0.00264)        & (0.00199)        & (0.00297)        \\
[1em]
meshnum       &                &                &                & -0.0253\sym{***} & -0.0244\sym{***} & -0.0258\sym{***} & -0.0283\sym{***} \\
              &                &                &                & (0.000187)       & (0.000453)       & (0.000261)       & (0.000345)       \\
[1em]
authorintl    &                &                &                & -0.0859\sym{***} & -0.0770\sym{***} & -0.0914\sym{***} & -0.103\sym{***}  \\
              &                &                &                & (0.00250)        & (0.00654)        & (0.00344)        & (0.00446)        \\
[1em]
Constant      & 0.510\sym{***} & 0.240\sym{***} & 0.266\sym{***} & 0.743\sym{***}   & 0.969\sym{***}   & 0.823\sym{***}   & -0.0185          \\
              & (0.00433)      & (0.0689)       & (0.0688)       & (0.0743)         & (0.111)          & (0.118)          & (0.198)          \\
\hline
Field fe      &                & $\checkmark$   & $\checkmark$   & $\checkmark$     & $\checkmark$     & $\checkmark$     & $\checkmark$     \\
Year fe       & $\checkmark$   &                & $\checkmark$   & $\checkmark$     & $\checkmark$     & $\checkmark$     & $\checkmark$     \\
Observations  & 5024732        & 5024732        & 5024732        & 4111637          & 1172896          & 1997524          & 941217           \\
\hline\hline
\multicolumn{7}{l}{\footnotesize Standard errors in parentheses} \\
\multicolumn{7}{l}{\footnotesize \sym{*} \(p<0.05\), \sym{**} \(p<0.01\), \sym{***} \(p<0.001\)} \\
\end{tabular}
\end{table*}

\begin{table*}[t]
\centering
\caption{Negative binomial regression modeling of dispersion of 10-year citations.}
\label{tab:c10v}
\begin{tabular}{l*{7}{c}}
\hline\hline
              & (1) & (2) & (3) & (4) & (5) & (6) & (7) \\
lnalpha       & 1980--2002 & 1980--2002 & 1980--2002 & 1980--2002 & 1980--1989 & 1990--1999 & 2000-2002 \\
\hline
ls            & 0.402\sym{***} & 0.314\sym{***} & 0.414\sym{***} & 0.262\sym{***}   & 0.276\sym{***}   & 0.235\sym{***}   & 0.303\sym{***}   \\
              & (0.00197)      & (0.00297)      & (0.00307)      & (0.00343)        & (0.00663)        & (0.00483)        & (0.00737)        \\
[1em]
ln\_authornum &                &                &                & -0.112\sym{***}  & -0.163\sym{***}  & -0.0864\sym{***} & -0.0884\sym{***} \\
              &                &                &                & (0.00117)        & (0.00218)        & (0.00169)        & (0.00251)        \\
[1em]
meshnum       &                &                &                & -0.0245\sym{***} & -0.0235\sym{***} & -0.0251\sym{***} & -0.0267\sym{***} \\
              &                &                &                & (0.000161)       & (0.000378)       & (0.000226)       & (0.000298)       \\
[1em]
authorintl=1  &                &                &                & -0.0667\sym{***} & -0.0571\sym{***} & -0.0696\sym{***} & -0.0879\sym{***} \\
              &                &                &                & 0.00221)         & (0.00571)        & (0.00306)        & (0.00392)        \\
[1em]
Constant      & 0.560\sym{***} & 0.369\sym{***} & 0.413\sym{***} & 0.862\sym{***}   & 1.066\sym{***}   & 0.859\sym{***}   & 0.234            \\
              & (0.00370)      & (0.0527)       & (0.0530)       & (0.0570)         & (0.0884)         & (0.0882)         & (0.135)          \\
\hline
Field fe      &                & $\checkmark$   & $\checkmark$   & $\checkmark$     & $\checkmark$     & $\checkmark$     & $\checkmark$     \\
Year fe       & $\checkmark$   &                & $\checkmark$   & $\checkmark$     & $\checkmark$     & $\checkmark$     & $\checkmark$     \\
Observations  & 5024732        & 5024732        & 5024732        & 4111637          & 1172896          & 1997524          & 941217           \\
\hline\hline
\multicolumn{7}{l}{\footnotesize Standard errors in parentheses} \\
\multicolumn{7}{l}{\footnotesize \sym{*} \(p<0.05\), \sym{**} \(p<0.01\), \sym{***} \(p<0.001\)} \\
\end{tabular}
\end{table*}

\begin{table*}[t]
\centering
\caption{Negative binomial regression modeling of dispersion of 15-year citations.}
\label{tab:c15v}
\begin{tabular}{l*{7}{c}}
\hline\hline
              & (1) & (2) & (3) & (4) & (5) & (6) & (7) \\
lnalpha       & 1980--2002 & 1980--2002 & 1980--2002 & 1980--2002 & 1980--1989 & 1990--1999 & 2000-2002 \\
\hline
ls            & 0.409\sym{***} & 0.329\sym{***} & 0.422\sym{***} & 0.271\sym{***}   & 0.282\sym{***}   & 0.249\sym{***}   & 0.314\sym{***}   \\
              & (0.00192)      & (0.00288)      & (0.00298)      & (0.00333)        & (0.00641)        & (0.00470)        & (0.00715)        \\
[1em]
ln\_authornum &                &                &                & -0.122\sym{***}  & -0.172\sym{***}  & -0.0952\sym{***} & -0.101\sym{***}  \\
              &                &                &                & (0.00114)        & (0.00212)        & (0.00164)        & (0.00243)        \\
[1em]
meshnum       &                &                &                & -0.0245\sym{***} & -0.0236\sym{***} & -0.0250\sym{***} & -0.0267\sym{***} \\
              &                &                &                & (0.000157)       & (0.000367)       & (0.000220)       & (0.000290)       \\
[1em]
authorintl    &                &                &                & -0.0614\sym{***} & -0.0503\sym{***} & -0.0637\sym{***} & -0.0813\sym{***} \\
              &                &                &                & (0.00216)        & (0.00559)        & (0.00300)        & (0.00382)        \\
[1em]
Constant      & 0.603\sym{***} & 0.404\sym{***} & 0.465\sym{***} & 0.913\sym{***}   & 1.101\sym{***}   & 0.846\sym{***}   & 0.377\sym{**}    \\
              & (0.00361)      & (0.0503)       & (0.0507)       & (0.0545)         & (0.0854)         & (0.0837)         & (0.129)          \\
\hline
Field fe      &                & $\checkmark$   & $\checkmark$   & $\checkmark$     & $\checkmark$     & $\checkmark$     & $\checkmark$     \\
Year fe       & $\checkmark$   &                & $\checkmark$   & $\checkmark$     & $\checkmark$     & $\checkmark$     & $\checkmark$     \\
Observations  & 5024732        & 5024732        & 5024732        & 4111637          & 1172896          & 1997524          & 941217           \\
\hline\hline
\multicolumn{7}{l}{\footnotesize Standard errors in parentheses}\\
\multicolumn{7}{l}{\footnotesize \sym{*} \(p<0.05\), \sym{**} \(p<0.01\), \sym{***} \(p<0.001\)}\\
\end{tabular}
\end{table*}

\begin{table*}[t]
\centering
\caption{Logistic regression modeling of top 1\% papers based on 3-year citations.}
\label{tab:h3}
\begin{tabular}{l*{7}{c}}
\hline\hline
              & (1) & (2) & (3) & (4) & (5) & (6) & (7) \\
h3            & 1980--2002 & 1980--2002 & 1980--2002 & 1980--2002 & 1980--1989 & 1990--1999 & 2000--2002 \\
\hline
ls            & -0.415\sym{***} & -0.963\sym{***} & -1.010\sym{***} & -0.754\sym{***} & -1.250\sym{***} & -0.695\sym{***} & -0.272\sym{***} \\
              & (0.0148)        & (0.0214)        & (0.0219)        & (0.0239)        & (0.0441)        & (0.0340)        & (0.0535)        \\
[1em]
ln\_authornum &                 &                 &                 & 0.962\sym{***}  & 0.957\sym{***}  & 0.961\sym{***}  & 0.987\sym{***}  \\
              &                 &                 &                 & (0.00947)       & (0.0169)        & (0.0136)        & (0.0214)        \\
[1em]
meshnum       &                 &                 &                 & 0.0826\sym{***} & 0.124\sym{***}  & 0.0840\sym{***} & 0.0599\sym{***} \\
              &                 &                 &                 & (0.00106)       & (0.00239)       & (0.00150)       & (0.00205)       \\
[1em]
authorintl=1  &                 &                 &                 & 0.299\sym{***}  & 0.183\sym{***}  & 0.330\sym{***}  & 0.326\sym{***}  \\
              &                 &                 &                 & (0.0130)        & (0.0308)        & (0.0177)        & (0.0244)        \\
[1em]
Constant      & -4.527\sym{***} & -4.240\sym{***} & -4.252\sym{***} & -5.956\sym{***} & -6.319\sym{***} & -6.320\sym{***} & -6.722\sym{***} \\
              & (0.0262)        & (0.0208)        & (0.0325)        & (0.0403)        & (0.0550)        & (0.0492)        & (0.0778)        \\
\hline
Field fe      &                 & $\checkmark$    & $\checkmark$    & $\checkmark$    & $\checkmark$    & $\checkmark$    & $\checkmark$    \\
Year fe       & $\checkmark$    &                 & $\checkmark$    & $\checkmark$    & $\checkmark$    & $\checkmark$    & $\checkmark$    \\
Observations  & 4985384         & 4985384         & 4985384         & 4080011         & 1159520         & 1984216         & 936193          \\
\hline\hline
\multicolumn{7}{l}{\footnotesize Standard errors in parentheses} \\
\multicolumn{7}{l}{\footnotesize \sym{*} \(p<0.05\), \sym{**} \(p<0.01\), \sym{***} \(p<0.001\)} \\
\end{tabular}
\end{table*}

\begin{table*}[t]
\centering
\caption{Logistic regression modeling of top 1\% papers based on 10-year citations.}
\label{tab:h10}
\begin{tabular}{l*{7}{c}}
\hline\hline
              & (1) & (2) & (3) & (4) & (5) & (6) & (7) \\
h10           & 1980--2002 & 1980--2002 & 1980--2002 & 1980--2002 & 1980--1989 & 1990--1999 & 2000--2002 \\
\hline
ls            & -0.205\sym{***} & -0.458\sym{***} & -0.480\sym{***} & -0.191\sym{***} & -0.871\sym{***} & -0.0625         & 0.405\sym{***}  \\
              & (0.0148)        & (0.0216)        & (0.0222)        & (0.0243)        & (0.0449)        & (0.0344)        & (0.0542)        \\
[1em]
ln\_authornum &                 &                 &                 & 0.818\sym{***}  & 0.805\sym{***}  & 0.842\sym{***}  & 0.786\sym{***}  \\
              &                 &                 &                 & (0.00940)       & (0.0167)        & (0.0135)        & (0.0212)        \\
[1em]
meshnum       &                 &                 &                 & 0.0803\sym{***} & 0.125\sym{***}  & 0.0809\sym{***} & 0.0570\sym{***} \\
              &                 &                 &                 & (0.00108)       & (0.00242)       & (0.00152)       & (0.00210)       \\
[1em]
authorintl=1  &                 &                 &                 & 0.280\sym{***}  & 0.192\sym{***}  & 0.296\sym{***}  & 0.321\sym{***}  \\
              &                 &                 &                 & (0.0134)        & (0.0319)        & (0.0183)        & (0.0251)        \\
[1em]
Constant      & -4.562\sym{***} &-4.422\sym{***} & -4.424\sym{***} & -5.937\sym{***} & -6.235\sym{***} & -6.346\sym{***} & -6.673\sym{***} \\
              & (0.0265)        & (0.0210)        & (0.0330)        & (0.0406)        & (0.0551)        & (0.0495)        & (0.0779)        \\
\hline
Field fe      &                 & $\checkmark$    & $\checkmark$    & $\checkmark$    & $\checkmark$    & $\checkmark$    & $\checkmark$    \\
Year fe       & $\checkmark$    &                 & $\checkmark$    & $\checkmark$    & $\checkmark$    & $\checkmark$    & $\checkmark$    \\
Observations  & 4985384         & 4985384         & 4985384         & 4080011         & 1159602         & 1984216         & 936193          \\
\hline\hline
\multicolumn{7}{l}{\footnotesize Standard errors in parentheses} \\
\multicolumn{7}{l}{\footnotesize \sym{*} \(p<0.05\), \sym{**} \(p<0.01\), \sym{***} \(p<0.001\)} \\
\end{tabular}
\end{table*}

\begin{table*}[t]
\centering
\caption{Logistic regression modeling of top 1\% papers based on 15-year citations.}
\label{tab:h15}
\begin{tabular}{l*{7}{c}}
\hline\hline
              & (1) & (2) & (3) & (4) & (5) & (6) & (7) \\
h15           & 1980--2002 & 1980--2002 & 1980--2002 & 1980--2002 & 1980--1989 & 1990--1999 & 2000--2002 \\
\hline
ls            & -0.0880\sym{***} & -0.203\sym{***} & -0.212\sym{***} & 0.0786\sym{**}  & -0.540\sym{***} & 0.207\sym{***}  & 0.595\sym{***}  \\
              & (0.0147)         & (0.0217)        & (0.0222)        & (0.0243)        & (0.0450)        & (0.0344)        & (0.0543)        \\
[1em]
ln\_authornum &                  &                 &                 & 0.733\sym{***}  & 0.735\sym{***}  & 0.760\sym{***}  & 0.663\sym{***}  \\
              &                  &                 &                 & (0.00931)       & (0.0165)        & (0.0134)        & (0.0209)        \\
[1em]
meshnum       &                  &                 &                 & 0.0765\sym{***} & 0.122\sym{***}  & 0.0759\sym{***} & 0.0535\sym{***} \\
              &                  &                 &                 & (0.00109)       & (0.00242)       & (0.00153)       & (0.00211)       \\
[1em]
authorintl=1  &                  &                 &                 & 0.297\sym{***}  & 0.209\sym{***}  & 0.323\sym{***}  & 0.322\sym{***}  \\
              &                  &                 &                 & (0.0135)        & (0.0321)        & (0.0184)        & (0.0253)        \\
[1em]
Constant      & -4.580\sym{***}  & -4.516\sym{***} & -4.520\sym{***} & -5.909\sym{***} & -6.237\sym{***} & -6.274\sym{***} & -6.524\sym{***} \\
              & (0.0267)         & (0.0212)        & (0.0331)        & (0.0409)        & (0.0552)        & (0.0494)        & (0.0775)        \\
\hline
Field fe      &                  & $\checkmark$    & $\checkmark$    & $\checkmark$    & $\checkmark$    & $\checkmark$    & $\checkmark$    \\
Year fe       & $\checkmark$     &                 & $\checkmark$    & $\checkmark$    & $\checkmark$    & $\checkmark$    & $\checkmark$    \\
Observations  & 4985384          & 4985384         & 4985384         & 4080011         & 1159602         & 1984216         & 936193          \\
\hline\hline
\multicolumn{7}{l}{\footnotesize Standard errors in parentheses}\\
\multicolumn{7}{l}{\footnotesize \sym{*} \(p<0.05\), \sym{**} \(p<0.01\), \sym{***} \(p<0.001\)}\\
\end{tabular}
\end{table*}

\end{document}